\documentclass[10pt,conference,letterpaper]{IEEEtran}
\IEEEoverridecommandlockouts
\usepackage{cite}
\usepackage{amsmath,amssymb,amsfonts}
\usepackage{algorithm,algpseudocodex}
\usepackage{graphicx}
\usepackage{textcomp}
\usepackage{xcolor}
\usepackage{multirow}
\usepackage{threeparttable}
\usepackage{booktabs}     
\usepackage{caption}
\usepackage{tabularx}  
\usepackage{dblfloatfix}
\usepackage{enumitem}
\usepackage{array}
\usepackage[utf8]{inputenc}

\newcommand{\insightbox}[1]{%
  \par\smallskip
  \begingroup
  \setlength{\fboxsep}{5pt}%
  \setlength{\fboxrule}{0.4pt}%
  \noindent\fcolorbox{black!25}{black!5}{%
    \parbox{\dimexpr\linewidth-2\fboxsep-2\fboxrule\relax}{#1}%
  }%
  \par
  \endgroup
  \smallskip
}

\def\BibTeX{{\rm B\kern-.05em{\sc i\kern-.025em b}\kern-.08em
    T\kern-.1667em\lower.7ex\hbox{E}\kern-.125emX}}
\begin{document}

\title{Beyond Host-Side Control: HBM Controller Partitioning Across Host and HBM Base Dies\\
}

\author{\IEEEauthorblockN{Tharunya Yogananda and Wantong Li}
\IEEEauthorblockA{
University of California, Riverside\\
Email: wantong.li@ucr.edu}
}

\maketitle

\begin{abstract}
Recent generations of high bandwidth memory (HBM), such as HBM4, allow the base die of the memory stack to be manufactured using a logic process that can accommodate more complex controller functions next to the DRAM. We present the first systematic quantitative study of function-level controller placement between the host and the HBM4 base die on a silicon interposer. We divide the controller into its main functions and evaluate four placements, from a complete host controller to a complete base die controller, with intermediate placements that move maintenance or scheduling to the base die. We extend Ramulator 2 with HBM4 timing and maintenance, model a die-to-die (D2D) link with timing checks at the DRAM interface, and compose area, energy, and temperature from synthesized controller blocks, modeled storage, and interface data. A provisioned D2D link preserves per-stack bandwidth and adds about twice its one-way delay to unloaded latency. Maintenance moves to the base die with no bandwidth cost and frees host command slots, and the freed share grows with stack height. Replacing the wide interface with a D2D link saves 61\% of host shoreline and 16.85 mm$^2$ of host area per stack. The D2D link dominates the added base die area and power, and the base die runs cooler under streaming with link energy below 0.4 pJ/bit. Shared control delivers up to 1.72$\times$ the throughput of static partitioning between two requestors.
\end{abstract}


\section{Introduction}

High bandwidth memory (HBM) supplies artificial intelligence (AI) accelerators with bandwidth through a wide interface that crosses a silicon interposer to the host processor. Each HBM stack places DRAM dies on a base die, and HBM4 widens the interface again while its base dies move to logic processes~\cite{jedec_hbm4}. Therefore, base dies manufactured in logic processes can accommodate more complex controller functions next to the DRAM. Some custom HBM designs already use this logic to replace the wide interface with a narrow die-to-die (D2D) link~\cite{marvell_chbm}.

In conventional HBM systems, the memory controller sits on the host die next to a wide physical interface (PHY), and each stack occupies host area and a segment of the host die edge, or shoreline, that compute could otherwise use. Moving controller functions into the stack reduces host logic and storage, while replacing the wide interface with D2D reduces PHY area and shoreline. These moves add logic, storage, and link power to the base die, and the D2D link adds delay to every access. Furthermore, stack height sharpens the trade-off. Taller stacks run hotter and hold more banks, so they refresh more often and carry more maintenance per channel~\cite{data_convection, chen_iedm25, jedec_hbm4}. Several compute dies may also share one stack, and a controller inside the stack can serve all of them. Industry disclosures report the outcome of custom designs without a function-level placement comparison, and no systematic study has separated which controller functions should move.

Prior work changes the memory attachment, the memory organization, or individual maintenance mechanisms. On-package memory over Universal Chiplet Interconnect Express (UCIe) attaches LPDDR and HBM to the host through a logic die and a standard narrow link~\cite{ucie_memory}, and the Hybrid Memory Cube places vault controllers in a logic layer behind a packetized serial interface~\cite{hmc}. These approaches establish alternative memory attachments and controller organizations. TokenStack adds processing in memory inside the HBM stack for language model inference~\cite{tokenstack}. Folded Banks~\cite{WLi-ISCA-25}, fine-grained DRAM~\cite{fgdram}, and RoMe~\cite{rome} reorganize banks, channels, or access granularity inside the stack. DreamRAM~\cite{dreamram} models the design space of custom 3D-stacked DRAM. DREAM~\cite{dream} and activation tracking designs~\cite{prac_tracker} improve RowHammer protection within a fixed controller location. None of these studies evaluates which controller functions should reside on the host or on the base die.

We present the first systematic quantitative study of function-level HBM4 controller placement between the host and the base die. We divide the controller into five components and evaluate four placements that move maintenance, scheduling, or the complete controller to the base die. We extend Ramulator 2~\cite{ramulator2} with HBM4 timing and maintenance, join it to a D2D link model with timing checks at the DRAM interface, and compose area, energy, and temperature from synthesized controller blocks, modeled storage, and interface data. The study contributes a partitioned HBM4 controller model, a placement analysis that covers bandwidth, latency, area, shoreline, energy, and temperature, and design rules for the shoreline and thermal break-even points of the D2D link. It shows that maintenance moves first at no bandwidth cost, that the D2D link dominates the added base die area and power, and that base die control integrates pooled access for multiple compute dies.

\section{HBM Controller Placement}
\label{sec:placement}

\begin{figure}[t]
\centering
\includegraphics[width=\columnwidth]{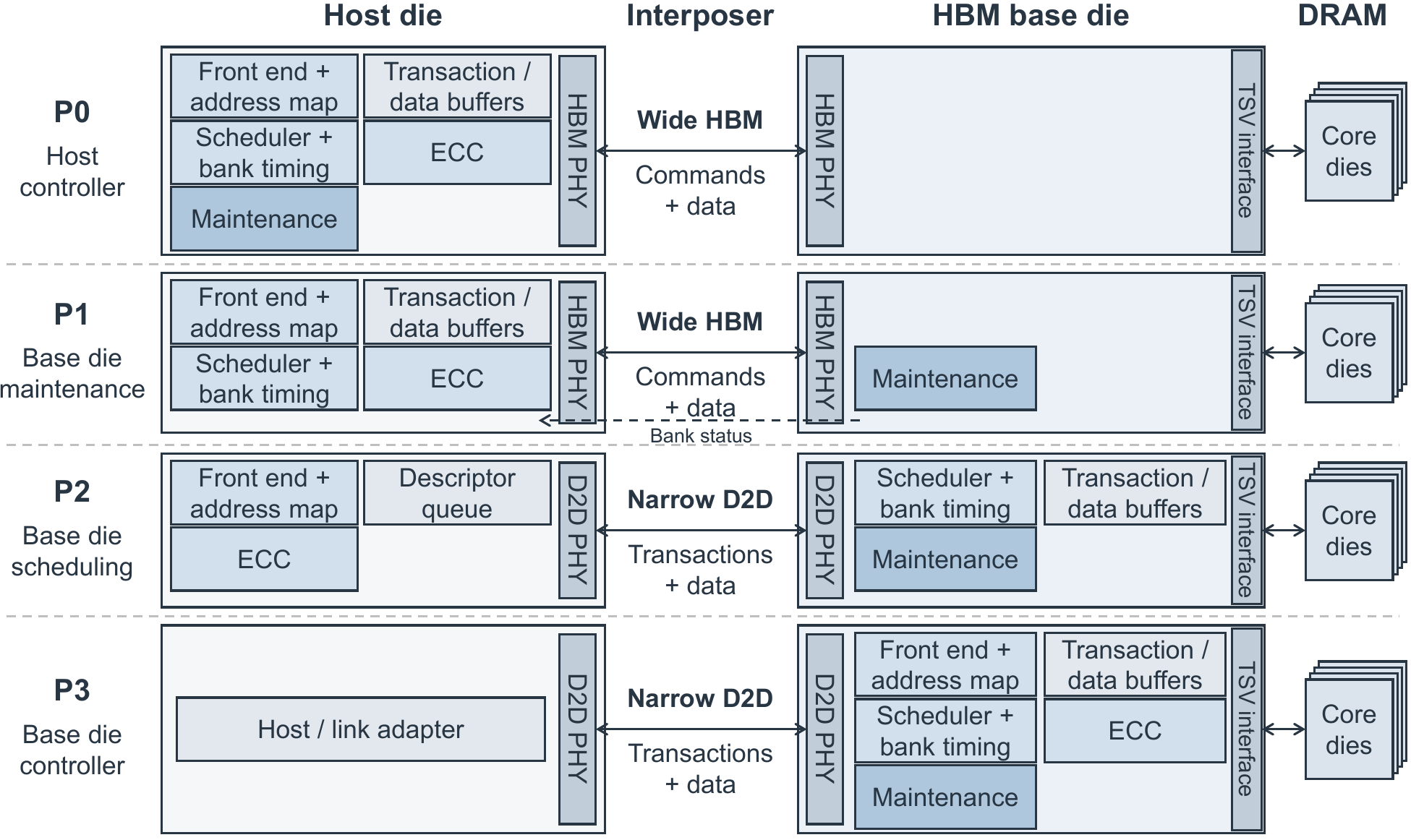}
\caption{Different HBM controller placements across the host die and base die.}
\label{fig:overview}
\vspace{-0.1in}
\end{figure}

An HBM stack places several DRAM core dies on a base die and connects them with through-silicon vias (TSVs). The stack sits next to the host processor on a silicon interposer, and each end of the interposer link has a PHY. HBM4 doubles the channel count of HBM3 to 32 channels, each split into two pseudo-channels (PCs), for 2,048 data wires at 8 Gb/s per wire~\cite{jedec_hbm4}. A recent 12-high HBM4 stack builds its base die in a 4 nm logic process and reaches 13 Gb/s per pin~\cite{samsung_isscc26}. The wide PHY on the host is large, and each attached stack consumes a long segment of the host shoreline.

The memory controller converts memory requests from the host into DRAM commands. It holds each request until the target bank is ready, orders commands to keep the data wires busy, and ensures that every command meets the DRAM timing rules. The controller also performs maintenance, which keeps the stored data correct over time. In current HBM systems, the complete controller sits on the host die next to its PHY.

We divide the controller into five components and consider several placements of them between the two dies. The front end receives requests and maps each physical address to a channel, PC, bank, row, and column. The transaction queues and data buffers hold pending requests together with their write and read data. The scheduler tracks the state of every bank, selects the next command, and enforces the timing rules, so it needs a current view of bank state. The maintenance engine issues refresh and RowHammer protection commands. The error correction code (ECC) logic encodes write data and checks read data. These components differ in the traffic they exchange with each other, in how much delay they tolerate, and in how much state they hold. The PHY terminates the interposer link and exists on both dies in every design. 

HBM4 adds work to several of these components. Each PC keeps its own queues and bank state. Refresh targets one bank or a whole PC, and the controller may postpone or pull in refresh commands within the limits of the standard~\cite{jedec_hbm4}. Refresh management (RFM) protects against RowHammer. The controller counts activations to each bank and issues an RFM command once the count reaches a threshold. Automatic error check and scrub (ECS) uses on-die ECC to scrub stored data during all-bank refresh or self-refresh, and the controller must issue all-bank refreshes often enough to meet the required scrub interval~\cite{jedec_hbm4}. The standard also shortens the refresh interval at high temperature, so a hotter stack spends more time in refresh~\cite{jedec_hbm4}. Taller stacks add stack IDs and banks to each channel, and the maintenance work per channel grows with stack height. The base die can host these functions next to the DRAM.

We study four placements of the controller components between the host and the base die, shown in Fig.~\ref{fig:overview}. The traffic that crosses the interposer defines each placement.

\noindent\textbf{Host controller} (P0): The host holds the complete controller and drives the wide interface. Every DRAM command crosses the interposer on the command wires, including activate, precharge, read, write, refresh, and RFM. Write data travels to the stack, and read data returns on the data wires.

\noindent\textbf{Base die maintenance} (P1): The wide interface remains, and refresh and RFM move to the base die. Only activate, precharge, read, and write commands cross the interposer together with the read and write data. The base die issues maintenance commands to the DRAM dies over a path inside the stack, and a status signal tells the host which banks are under maintenance.

\noindent\textbf{Base die scheduling} (P2): A narrow D2D link replaces the wide interface, as in recent custom HBM designs~\cite{marvell_chbm}. The host sends transactions, each carrying a read or write request with its address, together with the write data. The base die holds the scheduler, the data buffers, and the maintenance engine, and it converts each transaction into DRAM commands, so no DRAM command crosses the interposer. Read data and write acknowledgments return to the host, which keeps the front end, the address mapping, the ECC logic, and a descriptor queue.

\noindent\textbf{Base die controller} (P3): The interposer carries the same transactions and responses as in P2. The host keeps only a thin link adapter, and the base die holds the complete controller. Therefore, P2 and P3 share the same link traffic and differ in host and base die area and power.

In every placement, a PHY remains at both ends of the interposer link, and the TSV interface inside the stack stays unchanged. P1 keeps the wide interface and removes maintenance commands from the host. P2 and P3 replace the wide PHY pair with a D2D PHY pair, which trades host die area and shoreline for new logic, storage, and link power on the base die. P3 also lets several compute dies share one controller inside the stack.

\section{Evaluation}

\subsection{Experimental Setup}

\noindent\textbf{Performance modeling}. We model DRAM behavior with Ramulator 2~\cite{ramulator2} and extend its HBM4 model with corrected timing constraints, per-bank refresh with postponement and pull-in, activation counting with RFM, and maintenance issued from the base die. We match maintenance policies in placement-only comparisons. We simulate one 64-bit channel with two PCs and scale stack quantities to 32 channels. We connect the controller to a D2D link model with serialization, finite buffering, credits, and propagation delay. Under base die control (P2 and P3), each transaction crosses the link as a descriptor followed by its write data, and the link returns read data or a write acknowledgment. For P1, the host and the base die share one view of bank state, which represents the status signal. Every command that reaches the DRAM passes a timing check at the receiving interface. Table~\ref{tab:params} lists the modeling inputs.

\noindent\textbf{Physical modeling}. We implement the HBM controller logic in RTL and synthesize in a 45-nm library. We then project the area and power towards a 14-nm node~\cite{neurosim}. SRAM holds the transaction storage, the link buffers, and an activation-tracking table with a fixed per-stack capacity. The wide HBM4 PHY and D2D module areas come from bump-limited footprint estimates. The host shoreline saving is $S = 1 - k\rho_{\mathrm{wide}}/\rho_{\mathrm{D2D}}$, where $\rho$ is the bandwidth per millimeter of shoreline and $k$ is the D2D capacity summed over both directions and divided by the stack peak. $e_{\mathrm{D2D}}$ denotes link energy per wire bit. Link power multiplies the simulated wire traffic in both directions by $e_{\mathrm{D2D}}$, and the base die takes half of each interface energy. $\Delta P$ is the base die power change relative to P0. It adds the controller, SRAM, and D2D endpoint power and subtracts the base die share of the replaced wide interface. The peak DRAM temperature, $T_{\mathrm{peak}}$, equals the loaded P0 baseline plus $\Delta P$ times the thermal resistance from the base die to the hottest DRAM die. We sweep the baseline and the resistance over published ranges for each stack height~\cite{chen_iedm25,chiplet_thermal_review}. DRAM core power is held fixed in the baseline.

\noindent\textbf{Workloads and baselines}. The streaming workload reads sequential addresses across all PCs and banks, and the uniform random workload mixes 80\% reads with 20\% writes across the stack. Both run under sustained saturation at every stack height, and we discard warmup and measure sustained operation. Latency runs use low-load requests and short bursts on 4H and 16H stacks. A sweep with two requestors on a 4H stack raises the share of work from one requestor from 50\% to 100\%. The host controller on the wide interface (P0) is the reference. An additional baseline keeps scheduling on the host and sends each DRAM command over the same D2D link, with maintenance on the host to isolate the link and on the base die to isolate the scheduler location. For two requestors, we compare shared base die control with static PC partitioning and with a pooled host controller. The two workloads cover high and low row locality.

\begin{table}[t]
\centering
\caption{Modeling parameters.}
\label{tab:params}
\setlength{\tabcolsep}{3pt}
\begin{tabularx}{\columnwidth}{@{}>{\raggedright\arraybackslash}Xc@{}} 
\toprule
Parameter & Value \\
\midrule
Channels, PCs per channel & 32, 2 \\ 
Data rate per pin & 8 Gb/s \\ 
Stack height & 4H, 8H, 12H, 16H \\ 
Access size & 32 B \\ 
$t_{\mathrm{REFI}}$, $t_{\mathrm{RFCpb}}$ & 3.9 $\mu$s, 280 ns \\ 
RFM thresholds (initial, maximum) & 1024, 1536$^*$ \\ 
\midrule
Live transactions per channel & 128 \\ 
D2D wires per channel per direction & 16 \\ 
D2D rate per wire & 32 Gb/s~\cite{tsmc_isscc25} \\ 
D2D one-way latency, credits & 10 ns, 64 \\ 
Descriptor, command, header size & 16, 8, 8 B \\ 
\midrule
Logic area scaling (45-nm to 14-nm) & 0.209 \\ 
Wide PHY footprint & 8 $\times$ 2.5 mm~\cite{ucie_memory_slides} \\ 
D2D module footprint (64 lanes/direction) & 0.39 $\times$ 1.59 mm~\cite{ucie_memory_slides} \\ 
\midrule
Wide interface energy per useful bit & 0.9 pJ~\cite{ucie_memory_slides} \\ 
D2D energy, $e_{\mathrm{D2D}}$ (pJ/bit) & 0.6~\cite{tsmc_isscc25}, 0.3~\cite{ucie_memory_slides}, 0.19~\cite{nishi_jssc24} \\ 
Base die share of interface energy & 0.5$^*$ \\ 
Logic, SRAM power & 140 mW/mm$^2$, 50 mW/Mb$^*$ \\ 
Loaded P0 baseline, 16H & 85$^\circ$C$^\dagger$ \\ 
Thermal resistance, 16H & 2.33$^\circ$C/W$^\dagger$ \\ 
Refresh doubling, operating limit & 85, 95$^\circ$C~\cite{jedec_hbm4,amd_xapp1377} \\ 
\bottomrule
\multicolumn{2}{@{}l}{\footnotesize $^*$Assumed. $^\dagger$Nominal point from the literature.} \\ 
\end{tabularx} 
\end{table}


\subsection{Results and Discussion}
\label{sec:results}

\begin{table}[t]
\centering
\caption{Per-stack results for each placement.}
\label{tab:results}
\setlength{\tabcolsep}{3pt}
\begin{tabular}{lcccc}
\toprule
Metric & P0 & P1 & P2 & P3 \\
\midrule
Host area saved (mm$^2$) & 0 & 0.24 & 15.58 & 16.85 \\
Host shoreline saved & 0 & 0 & 61\% & 61\% \\
Interposer data wires & 2,048 & 2,048 & 1,024 & 1,024 \\
Total base area added (mm$^2$) & 0 & 0.24 & 5.85 & 7.06 \\
Base controller power (W) & 0 & 0.12 & 0.32 & 0.49 \\
$\Delta P$ (W)$^*$ & 0 & 0.12 & 1.77 & 1.94 \\
$T_{\mathrm{peak}}$ ($^\circ$C)$^*$ & 85.0 & 85.3 & 89.1 & 89.5 \\
Added latency (ns)$^\dagger$ & 0 & 0 & 20.1--22.8 & 20.1--22.8 \\
\bottomrule
\end{tabular}

\vspace{2pt}
\parbox{\columnwidth}{\footnotesize $^*$16H, streaming, $e_{\mathrm{D2D}} = 0.6$ pJ/bit, 85$^\circ$C baseline, 2.33$^\circ$C/W. $^\dagger$Unloaded, 10 ns one-way link.}
\end{table}

\noindent\textbf{Maintenance placement}. Moving refresh and RFM to the base die removes their commands from the host command wires. The removed occupancy grows with stack height, from 0.4\% of command slots at 4H to 1.63\% at 16H, because taller stacks add banks that each need refresh. With the same maintenance policy on both dies, P1 delivers 0.10\% to 0.18\% more bandwidth than P0 on random traffic at 16H. The maintenance engine and its tracking state add 0.24 mm$^2$ and 0.12 W to the base die, which raises $T_{\mathrm{peak}}$ by 0.28$^\circ$C at 16H.

\insightbox{%
\noindent\textbf{Insight 1}. Maintenance is the first function to move. It costs no bandwidth and little area or heat, and the share of host command slots it frees grows with stack height.
}

\begin{figure}[t]
\centering
\includegraphics[width=\columnwidth]{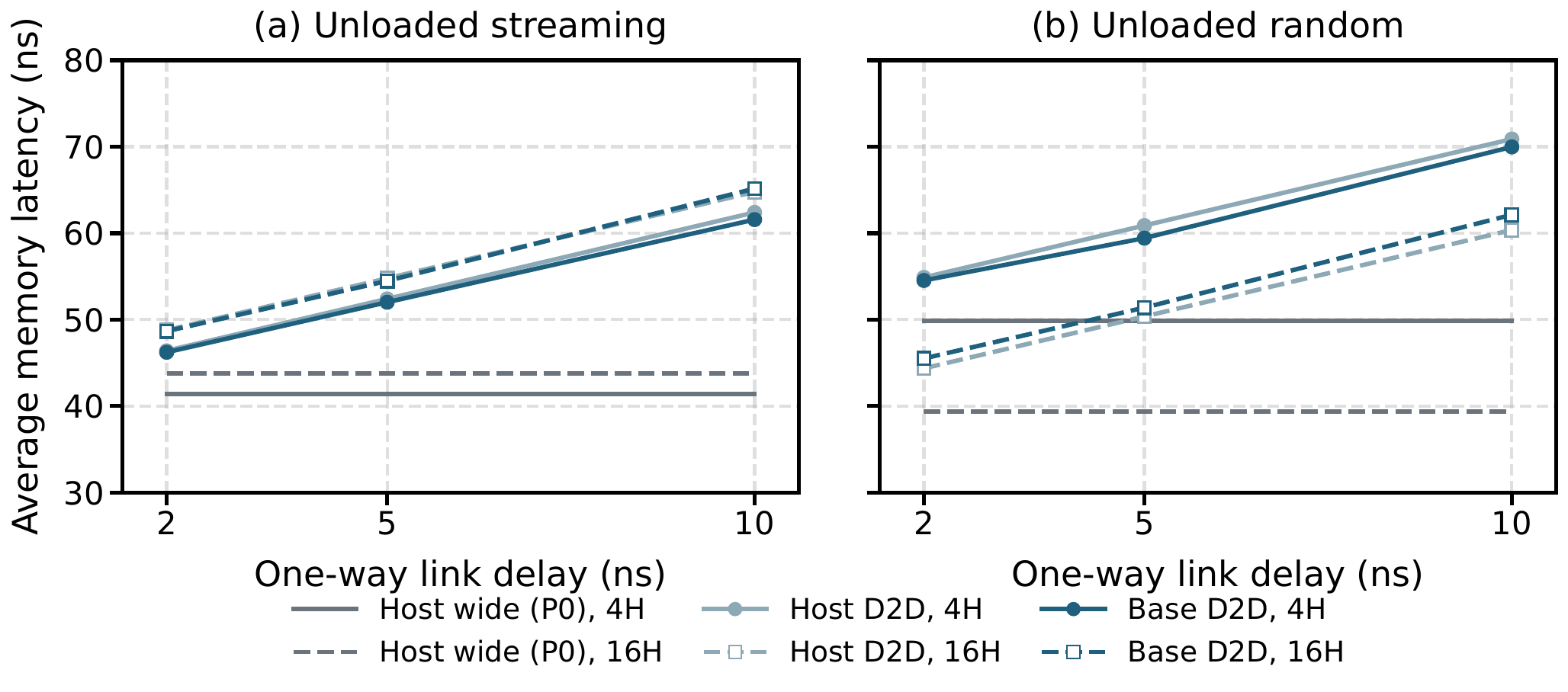}
\caption{Absolute average latency for (a) unloaded streaming and (b) random traffic. Host wide denotes P0, host D2D retains host scheduling, and base D2D represents P2/P3.}
\label{fig:bw}
\vspace{-0.1in}
\end{figure}

\noindent\textbf{Bandwidth and latency}. Fig.~\ref{fig:bw} compares absolute average memory latency for unloaded streaming and random traffic on 4H and 16H stacks. With 16 wires per channel in each direction at 32 Gb/s, a 10 ns link, and 64 credits, host control over the D2D link delivers 0.9977$\times$ to 1.0006$\times$ the bandwidth of P0 on both workloads. This baseline isolates the link. With maintenance on the base die in both cases, P2 and P3 deliver 0.9949$\times$ to 1.0014$\times$ the bandwidth of host scheduling over the same link. A link passes when it keeps at least 99\% of the baseline bandwidth. The narrowest passing link uses 14 wires per direction at 4H, 8H, and 12H and 12 wires at 16H, where the stack delivers less bandwidth. A link of 21 wires at 25.2 Gb/s~\cite{nishi_jssc24} also preserves bandwidth at 4H and 16H. The link adds about twice its one-way delay to unloaded latency, 20.1 to 22.8 ns at 10 ns. At 40 ns, longer flight time meets the 128-transaction cap and holds streaming to 0.84$\times$ to 0.93$\times$ of P0, and doubling the credits from 128 to 256 does not recover it. The TSV interface inside the stack is identical in every placement, so the vertical bandwidth of the stack does not change.

\insightbox{%
\noindent\textbf{Insight 2}. Controller placement preserves bandwidth with sufficient link capacity and transaction storage. Longer links increase the storage needed to sustain throughput.
}

\begin{figure}[t]
\centering
\includegraphics[width=\columnwidth]{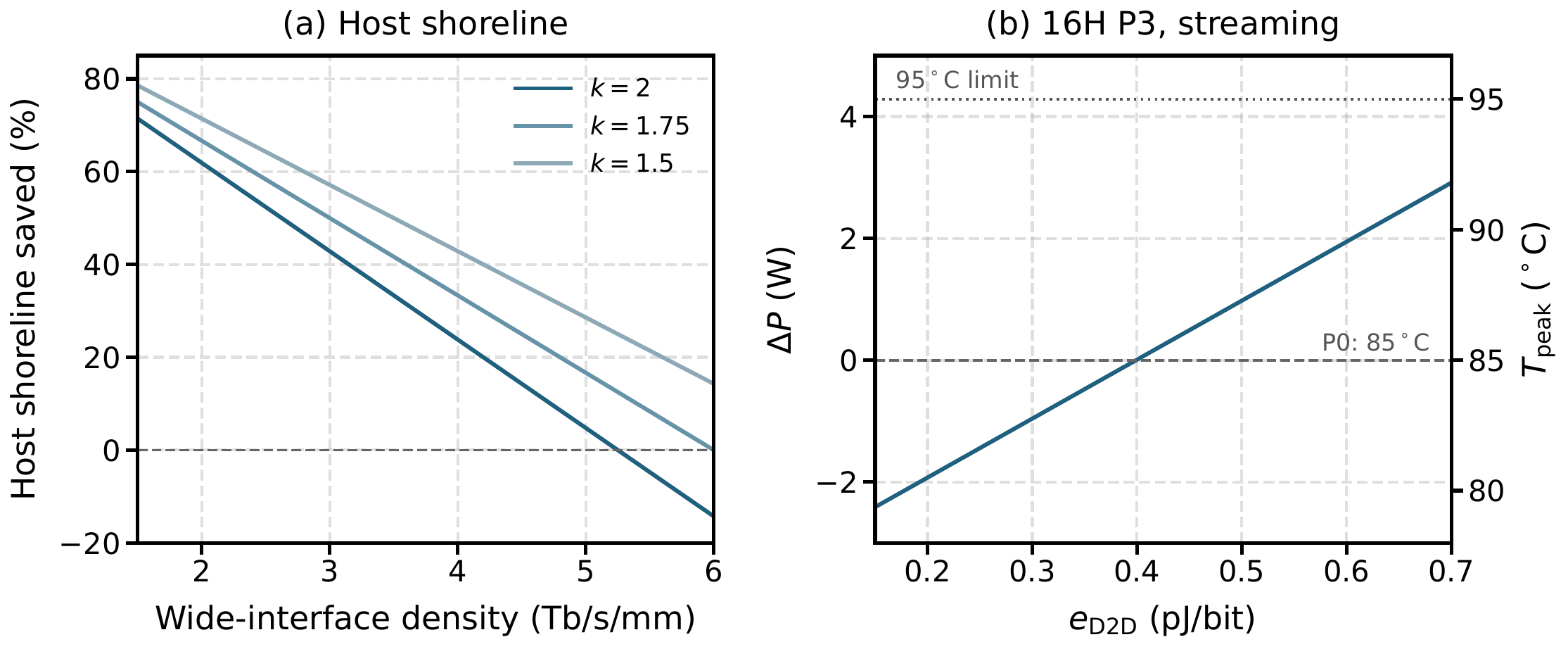}
\caption{(a) Shoreline density sets the host shoreline saving for link capacities $k = 2$, $1.75$, and $1.5$. (b) $e_{\mathrm{D2D}}$ sets $\Delta P$ and $T_{\mathrm{peak}}$ of P3 for 16H streaming.}
\label{fig:phys}
\vspace{-0.1in}
\end{figure}

\noindent\textbf{Host die}. Replacing the wide interface with the D2D link frees most of the host shoreline. The wide PHY carries 2.05 Tb/s per millimeter of shoreline, while the D2D PHY carries 10.5 Tb/s per millimeter summed over both directions~\cite{tsmc_isscc25}. With full capacity in each direction ($k = 2$), the link saves 61\% of the host shoreline, and the 14-wire and 12-wire links raise the saving to 66\% and 71\% respectively. Fig.~\ref{fig:phys}(a) shows the saving across shoreline densities. The link saves shoreline whenever its density exceeds $k$ times the wide density, so the break-even wide density is 5.25 Tb/s per millimeter at $k = 2$. A 13 Gb/s DRAM with the same wide PHY footprint and a proportionally wider link saves 37\%. The link also halves the data wires on the interposer, from 2,048 to 1,024 lanes, while each lane runs four times faster. P3 saves 16.85 mm$^2$ of host area and P2 saves 15.58 mm$^2$. With a smaller HBM4 PHY area of 15 mm$^2$~\cite{guc_phy_area}, P3 saves 11.85 mm$^2$. Of the P3 saving, 15.07 mm$^2$ comes from replacing the wide PHY and 1.79 mm$^2$ from the net reduction in controller logic and SRAM. P2 and P3 send transactions and use 26\% fewer wire bits than host control over the same link on random traffic, where row misses require additional activate and precharge commands. On streaming traffic, they use 12.5\% more wire bits.

\insightbox{%
\noindent\textbf{Insight 3}. Replacing the wide PHY supplies most of the host saving, and moving controller logic adds the rest. Shoreline savings depend on link density and provisioned capacity.
}

\noindent\textbf{Base die}. The following energy and thermal results use 16H streaming. P3 adds 7.06 mm$^2$ of total area to the base die and P2 adds 5.85 mm$^2$, without subtracting the wide PHY that the base die no longer needs. The D2D PHY accounts for 4.93 mm$^2$ of each. The P3 controller and its SRAM draw 0.49 W. The D2D endpoint accounts for 92\% of endpoint-plus-controller power at $e_{\mathrm{D2D}} = 0.6$ pJ/bit. Packet headers and read descriptors double the wire traffic relative to useful data. The D2D interface spends 1.2 pJ, 0.6 pJ, and 0.38 pJ per useful bit at the three link energies, against 0.9 pJ for the wide interface. Therefore, the D2D interface saves energy only when $e_{\mathrm{D2D}} < 0.45$ pJ/bit. Fig.~\ref{fig:phys}(b) shows $\Delta P$ and $T_{\mathrm{peak}}$ at 16H. Under the nominal thermal assumptions, with the measured $e_{\mathrm{D2D}} = 0.6$ pJ/bit, $\Delta P$ is 1.94 W and $T_{\mathrm{peak}}$ reaches 89.5$^\circ$C, within the same refresh tier as P0 and below the 95$^\circ$C limit. With the UCIe target of $e_{\mathrm{D2D}} = 0.3$ pJ/bit, $\Delta P$ falls to $-0.96$ W and the stack cools to 82.8$^\circ$C. $\Delta P$ reaches zero at $e_{\mathrm{D2D}} = 0.4$ pJ/bit.

\insightbox{%
\noindent\textbf{Insight 4}. The D2D link dominates the added area and power, while controller logic contributes a small share. Under fixed cooling, link energy determines whether the base die runs cooler or hotter than P0.
}

\begin{figure}[t]
\centering
\includegraphics[width=0.65\columnwidth]{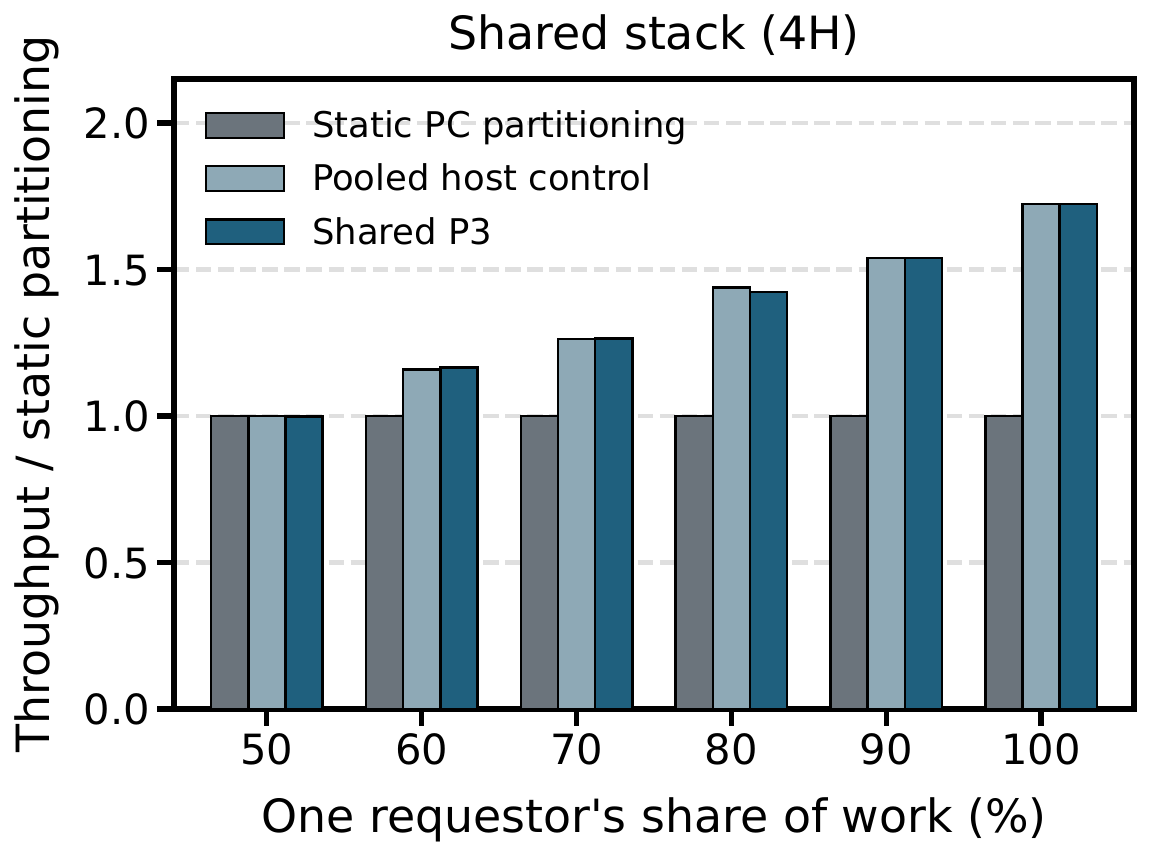}
\caption{Pooling recovers stranded bandwidth under uneven requestor demand.}
\label{fig:share}
\vspace{-0.1in}
\end{figure}

\noindent\textbf{Shared stack}. Fig.~\ref{fig:share} compares two requestors that share one stack. Static partitioning gives each requestor one PC. At balanced load, shared P3 matches it at 0.998$\times$. As one requestor takes a larger share of the work, the static partition leaves the other PC idle, and shared control reaches 1.72× when one requestor does all the work. The pooled host matches shared P3 within 1.1\%. Therefore, the gain comes from pooling channel capacity. 
When each of two compute dies would otherwise carry a full controller and PHY, a single P3 controller saves 33.7 mm$^2$ of host area and 9.8 mm of host shoreline in total, and adds 7.06 mm$^2$ to the base die once.

\insightbox{%
\noindent\textbf{Insight 5}. Pooling channel capacity recovers the throughput that static partitioning strands. Placing the shared controller on the base die enables pooling without replicating a controller and wide PHY on each compute die.
}

\section{Conclusion}
We study where the functions of an HBM4 controller should reside between the host and the base die. Controller placement preserves per-stack bandwidth when each channel has a channel-sized link, and longer links need more outstanding transactions. Maintenance is the first function to move, since it costs no bandwidth and frees host command slots as stacks grow taller. Moving the scheduler and the complete controller with a narrow D2D link saves most of the host shoreline and the area of the wide PHY. The D2D link then dominates the added area and power. An efficient D2D link can reduce base die power and temperature. When several compute dies share one stack, pooling recovers stranded channel capacity, and P3 reduces the controller and PHY area that each compute die would otherwise replicate.



\bibliographystyle{IEEEtran}
\bibliography{citations}

@ARTICLE{WLi-ISCA-25,
  author={Adhinarayanan, Vignesh and Beckmann, Brad and Li, Wantong and Seyedzadeh, Mohammad and Blagodurov, Sergey and Aguren, Derrick and Lee, Hayden},
  journal={ACM/IEEE  International Symposium on Computer Architecture (ISCA)}, 
  title={Folded Banks: 3D-stacked HBM design for fine-grained random-access bandwidth}, 
  year={2025},
  volume={},
  number={},
  pages={},
  doi={}}

@techreport{jedec_hbm4,
  author = {{JEDEC Solid State Technology Association}},
  title = {High Bandwidth Memory ({HBM4}) {DRAM}},
  institution = {JEDEC Solid State Technology Association},
  number = {JESD270-4},
  year = {2025},
  month = apr
}

@misc{marvell_chbm,
  author = {{Marvell Technology}},
  title = {Marvell Announces Breakthrough Custom {HBM} Compute Architecture to Optimize Cloud {AI} Accelerators},
  year = {2024},
  month = dec,
  howpublished = {Press release},
  url = {https://www.marvell.com/company/newsroom/marvell-announces-breakthrough-custom-hbm-compute-architecture.html}
}

@article{data_convection,
  author = {Soheil Khadirsharbiyani and Jagadish Kotra and Karthik Rao and Mahmut Kandemir},
  title = {Data Convection: A {GPU}-Driven Case Study for Thermal-Aware Data Placement in {3D DRAMs}},
  journal = {Proceedings of the ACM on Measurement and Analysis of Computing Systems},
  volume = {6},
  number = {1},
  year = {2022},
  month = mar,
  articleno = {7},
  doi = {10.1145/3508027}
}

@inproceedings{chen_iedm25,
  author = {Yukai Chen and Melina Lofrano and Diksha Moolchandani and Herman Oprins and Geert Van der Plas and Julien Ryckaert and Dwaipayan Biswas and James Myers},
  title = {Breaking Thermal Bottleneck in {3D HBM}-on-{GPU} Integration via System-Technology Co-Optimization},
  booktitle = {2025 IEEE International Electron Devices Meeting (IEDM)},
  year = {2025},
  pages = {1--4},
  doi = {10.1109/IEDM50572.2025.11353711}
}

@inproceedings{ucie_memory,
  author = {Das Sharma, Debendra and Choudhary, Swadesh and Onufryk, Peter and Pelt, Rob},
  title = {On-Package Memory with Universal Chiplet Interconnect Express ({UCIe}): A Low Power, High Bandwidth, Low Latency and Low Cost Approach},
  booktitle = {2025 IEEE Symposium on High-Performance Interconnects (HOTI)},
  year = {2025},
  pages = {87--96}
}

@inproceedings{hmc,
  author = {J. Thomas Pawlowski},
  title = {Hybrid Memory Cube ({HMC})},
  booktitle = {2011 IEEE Hot Chips 23 Symposium (HCS)},
  year = {2011},
  pages = {1--24},
  doi = {10.1109/HOTCHIPS.2011.7477494}
}

@article{tokenstack,
  title={TokenStack: A Heterogeneous HBM-PIM Architecture and Runtime for Efficient LLM Inference},
  author={Li, Zhuoran and Bian, Zhuohang and Huang, Zihao and Zhao, Yibo and Li, Xueqi and Sun, Guangyu and Zhuo, Youwei},
  journal={arXiv preprint arXiv:2605.05639},
  year={2026}
}

@inproceedings{fgdram,
  author = {Mike O'Connor and Niladrish Chatterjee and Donghyuk Lee and John Wilson and Aditya Agrawal and Stephen W. Keckler and William J. Dally},
  title = {Fine-Grained DRAM: Energy-Efficient {DRAM} for Extreme Bandwidth Systems},
  booktitle = {Proceedings of the 50th Annual IEEE/ACM International Symposium on Microarchitecture (MICRO)},
  year = {2017},
  doi = {10.1145/3123939.3124545},
}

@inproceedings{rome,
  author = {Hwayong Nam and Seungmin Baek and Jumin Kim and Michael Jaemin Kim and Jung Ho Ahn},
  title = {{RoMe}: Row Granularity Access Memory System for Large Language Models},
  booktitle = {2026 IEEE International Symposium on High Performance Computer Architecture (HPCA)},
  year = {2026},
  pages = {1--15}
}

@inproceedings{dreamram,
  title={Dreamram: A fine-grained configurable design space modeling tool for custom 3d die-stacked dram},
  author={Cai, Victor and Zhou, Jennifer and Brooks, David and Wei, Gu-Yeon},
  booktitle={2026 Design, Automation \& Test in Europe Conference (DATE)},
  pages={1--7},
  year={2026},
  organization={IEEE}
}

@inproceedings{dream,
  author = {Hritvik Taneja and Moin Qureshi},
  title = {{DREAM}: Enabling Low-Overhead Rowhammer Mitigation via Directed Refresh Management},
  booktitle = {Proceedings of the 52nd Annual International Symposium on Computer Architecture (ISCA)},
  year = {2025},
  pages = {776--792},
  doi = {10.1145/3695053.3731117}
}

@inproceedings{prac_tracker,
  author = {Moinuddin Qureshi and Salman Qazi},
  title = {{MOAT}: Securely Mitigating Rowhammer with Per-Row Activation Counters},
  booktitle = {Proceedings of the 30th ACM International Conference on Architectural Support for Programming Languages and Operating Systems, Volume 1 (ASPLOS)},
  year = {2025},
  doi = {10.1145/3669940.3707278}
}

@article{ramulator2,
  author = {Haocong Luo and Yahya Can Tu\u{g}rul and F. Nisa Bostanc{\i} and Ataberk Olgun and A. Giray Ya\u{g}l{\i}k\c{c}{\i} and Onur Mutlu},
  title = {{Ramulator 2.0}: A Modern, Modular, and Extensible {DRAM} Simulator},
  journal = {IEEE Computer Architecture Letters},
  volume = {23},
  number = {1},
  pages = {112--116},
  year = {2024},
  doi = {10.1109/LCA.2023.3333759}
}

@inproceedings{samsung_isscc26,
  author = {Sunghwan Joo and Jinyeon Kim and Yongsun Lee and Ji-Young Kim and Youngsik Lee and Yong-Min Kim and ChiSung Oh and Kyu-Ha Shim and Haesuk Lee and Young-Yong Byun and ChangHyun Bae and Joohwan Kim and Je-Min Ryu and Shin-Haeng Kang and Jaehoon Lee and Young-Uk Chang and JaeKyung Lee and JongTae Hwang and Daehwan Seo and Ki-Heon Na and Young Guen Song and Daihyun Lim and Kyung-Soo Ha and Young-Soo Sohn and Sang-Joon Hwang},
  title = {A 36GB 3.3TB/s {HBM4 DRAM} with Per-Channel {TSV RDQS} Auto Calibration and Fully-Programmable {MBIST}},
  booktitle = {2026 IEEE International Solid-State Circuits Conference (ISSCC)},
  year = {2026},
  pages = {264--266},
  doi = {10.1109/ISSCC49663.2026.11409139}
}

@article{neurosim,
  author = {Junmo Lee and Anni Lu and Wantong Li and Shimeng Yu},
  title = {NeuroSim V1.4: Extending Technology Support for Digital Compute-in-Memory Toward 1nm Node},
  journal = {IEEE Transactions on Circuits and Systems I: Regular Papers},
  year = {2024},
  volume = {71},
  number = {4},
  pages = {1733--1744},
  doi = {10.1109/TCSI.2024.3362822}
}

@misc{ucie_memory_slides,
  author = {Das Sharma, Debendra and Choudhary, Swadesh and Onufryk, Peter and Pelt, Rob},
  title = {On-Package Memory with Universal Chiplet Interconnect Express (UCIe): A Low-Power, High-Bandwidth, Low-Latency, and Low-Cost Approach},
  howpublished = {HOTI 2025 presentation},
  year = {2025},
  month = aug
}

@inproceedings{tsmc_isscc25,
  author = {Mu-Shan Lin and Chien-Chun Tsai and Shenggao Li and Wei-Chih Chen and Wen-Hung Huang and Yu-Chi Chen and Yu-Jie Huang and Alan Drake and Chin-Hua Wen and Paul Ranucci and Hsin-Hung Kuo and Aidong Yin and Shu-Chun Yang and Farsheed Mahmoudi and Han-Tzung Ke and Chao-Chieh Li and Nai-Chen Cheng and Jimmy Wang and Kevin Lin and Harry Liao and Jie-Ren Huang and Meng-Hsuan Wu and Kenny Cheng-Hsiang Hsieh and Nicholas Amatruda and William Polanco and David King and Todd Basso and Anwar Kashem},
  title = {36.1 A 32Gb/s 10.5Tb/s/mm 0.6pJ/b {UCIe}-Compliant Low-Latency Interface in 3nm Featuring Matched-Delay for Dynamic Clock Gating},
  booktitle = {2025 IEEE International Solid-State Circuits Conference (ISSCC)},
  year = {2025},
  pages = {586--588},
  doi = {10.1109/ISSCC49661.2025.10904767}
}

@article{nishi_jssc24,
  author = {Yoshinori Nishi and John W. Poulton and Walker J. Turner and Xi Chen and Sanquan Song and Brian Zimmer and Stephen G. Tell and Nikola Nedovic and John M. Wilson and William J. Dally and C. Thomas Gray},
  title = {A 0.190-pJ/bit 25.2-Gb/s/wire Inverter-Based {AC}-Coupled Transceiver for Short-Reach Die-to-Die Interfaces in 5-nm {CMOS}},
  journal = {IEEE Journal of Solid-State Circuits},
  year = {2024},
  volume = {59},
  number = {4},
  pages = {1146--1157},
  doi = {10.1109/JSSC.2023.3338478}
}

@article{chiplet_thermal_review,
  author = {Darpan Virmani and Baibhab Chatterjee},
  title = {Thermal Management Challenges in {2.5D} and {3D} Chiplet Integration: A Review on Architecture--Cooling Co-Design},
  journal = {Eng},
  year = {2025},
  volume = {6},
  number = {12},
  pages = {373},
  doi = {10.3390/eng6120373},
  url = {https://www.mdpi.com/2673-4117/6/12/373}
}

@misc{guc_phy_area,
  author = {Anton Shilov},
  title = {{HBM} Undergoes Major Architectural Shakeup as {TSMC} and {GUC} Detail {HBM4}, {HBM4E} and {C-HBM4E}},
  howpublished = {Tom's Hardware},
  year = {2025},
  month = dec,
  url = {https://www.tomshardware.com/pc-components/dram/hbm-undergoes-major-architectural-shakeup-as-tsmc-and-guc-detail-hbm4-hbm4e-and-c-hbm4e-3nm-base-dies-to-enable-2-5x-performance-boost-with-speeds-of-up-to-12-8gt-s-by-2027},
  note = {Dec. 2, 2025}
}

@techreport{amd_xapp1377,
  author = {{Advanced Micro Devices}},
  title = {Designing Thermal Solutions for {AMD} Embedded Devices},
  institution = {Advanced Micro Devices},
  number = {XAPP1377},
  year = {2025},
  month = jun,
  note = {Revision 2.0, June 27, 2025},
  url = {https://docs.amd.com/r/en-US/xapp1377-heatsinks-thermal/Obtaining-Thermal-and-Power-Targets-to-Use-with-Thermal-Simulation}
}

\end{document}